# Magneto-transport properties of oriented $Mn_2CoAl$ films sputtered on thermally oxidized Si substrates

G. Z. Xu,[1] Y. Du,[1] X. M. Zhang,[1] H. G. Zhang,[2] E. K. Liu,[1] W. H. Wang[1,a)] and G. H. Wu[1]

[1]State Key Laboratory for Magnetism, Beijing National Laboratory for Condensed Matter Physics, Institute of Physics, Chinese Academy of Sciences, Beijing 100190, People's Republic of China

[2]College of Materials Science and Engineering, Beijing University of Technology, Beijing 100124, People's Republic of China

Spin gapless semiconductors are interesting novel class of materials by embracing both magnetism and semiconducting. Its potential application in future spintronics requires realization in thin film form. In this letter, we report a successful growth of spin gapless $Mn_2CoAl$ films on thermally oxidized Si substrates by magnetron sputtering deposition. The films deposited at 673K are well oriented to (001) direction and display a uniform-crystalline surface. Magnetotransport measurements on the oriented films reveal a semiconducting-like resistivity, small anomalous Hall conductivity and linear magnetoresistance (MR) representative of the transport signatures of spin gapless semiconductors. The magnetic properties of the films have also been investigated and compared to that of bulk $Mn_2CoAl$, with small discrepancy induced by the composition deviation.

------------------------------

a) Author to whom correspondence should be addressed. Electronic mail: wenhong.wang@iphy.ac.cn

Spin gapless semiconductors (SGS) are a new class of spintronic materials, which is characterized by the existence of semiconducting gap in one spin channel and a zero gap in the other across the Fermi level[1, 2]. The unique band characteristics of SGS indicate novel transport properties such as fully spin polarized carriers, zero excitation energy and external field sensibility, making them particularly attractive for future spintronic applications. Since the first theoretical prediction of SGS, several Pd-based oxide gapless semiconductors with doping by Co and Mn have demonstrated experimentally to show SGS-like transport properties, such as linear magnetoresistance, low Hall conductivity and high mobility. More recently, several theoretical works have predicted new candidates of SGS in modified graphene[3] and Heusler compounds[4-7]. Among the Heusler candidates, $Mn_2CoAl$ was especially attracted due to its unique spin gapless transport properties having been recently revealed in bulk form[8]. Development of spintronic devices utilizing SGS will require accurate characterization of the magnetic and transport properties of thin films. Nevertheless, the growth of $Mn_2CoAl$ films has, so far, only been reported on GaAs(001) substrates using molecular-beam epitaxy (MBE) exhibiting a metallic-like resistivity at low temperatures[9]. In this letter, we report the growth of oriented $Mn_2CoAl$ films on thermally oxidized Si substrates using magnetron sputtering. Magnetotransport measurements on the oriented films revealed a semiconducting-like resistivity, small Hall conductivity and linear magnetoresistance indicative of the transport features of SGS.

We have deposited a series of $Mn_2CoAl$ films on thermally oxidized Si substrates using magnetron sputtering system with the base pressure of below $5\times10^{-5}$ Pa. The optimized sample we chose for present study is deposited under an Ar pressure of 0.2Pa while the substrate was kept at 673K with dc power of 30W. The typical thickness of the film determined by the stylus profile is about 200-nm-thick under a deposition rate of ~4 Å/s. The structure was determined by X-ray diffraction (XRD) using the Cu-Kα radiation. The composition and morphology of the film were examined by scanning electron microscope (SEM) equipped with an energy-dispersive X-ray spectroscopy (EDX). The magnetic and transport measurements were performed using a superconducting quantum interference device magnetometer (MPMS, Quantum Design) and Physics property measurement system (PPMS, Quantum Design), respectively. The electronic band structure is calculated by employing the coherent potential approximation (CPA)[10] for nonstoichiometric situation.

Figure 1 shows X-ray $\theta$-$2\theta$ diffraction pattern of 200-nm-thick $Mn_2CoAl$ film (blue line) along with the oxidized Si substrate (grey line). It can be found that, besides the peaks from Si substrate, the peaks are indexed to (002) and (004) diffraction direction comparing to bulk $Mn_2CoAl$, displaying oriented texture growth even in this amorphous substrate. Observation of the superlattice (002) diffraction implies good order of the film. The lattice constant is determined to be 5.80 Å, nearly equal to the bulk. The SEM image in the inset shows homogenous crystal grain growth, and the composition determined by the EDX is $Mn_2Co_{0.7}Al_{0.9}$, with Al in excess. The measurement uncertainty of 5% was determined by repeating measurements on the same sample.

Figure 2(a) shows the temperature dependence of the electrical resistivity $\rho_{xx}(T)$ in the range from 5K to 300K. The increase in resistivity with decreasing temperature is a clear signature that the oriented $Mn_2CoAl$ film exhibits a semiconducting-like behavior in the whole region, which is different from the $Mn_2CoAl$ films fabricated by MBE with a turning point at low temperatures.[9] Actually, The sign change of $\rho_{xx}(T)$ slope is commonly observed in semimetals or narrow-gap semiconductors where atomic disorder, defects or nonstoichiometry gives rise to impurity levels governing low temperature conduction.[11,12] In our case, the general shape of $\rho_{xx}(T)$ is characteristics of semimetals or doped semiconductors which can be described by an thermal activation model with a small energy gap.[13] As shown in the inset of Fig. 2(a), the conductivity is well approximated by a simple model:[11] $\sigma(T) = 1/\rho(T) = \sigma_0 + \sigma_a e^{-E_g/k_B T}$ , where $\sigma_0$ indicates possible nonstoichiometry generated conduction at low temperatures. The best parameter fit yields an energy gap of $E_g = 90\text{meV}$.

The magnetoresistance (MR) in Fig. 2(b) manifest linear trend up to 5T field, similar to the bulk $Mn_2CoAl$ case and other gapless systems.[8, 14] But comparing to the bulk $Mn_2CoAl$, the MR magnitude of the oriented film becomes smaller and no sign change occurs in the entire temperature range studied. The positive MR at low temperatures was not observed probably due to the Fermi-level shift with the composition deviation. Nevertheless, the MR presents a maximum around 50K (as seen in the inset) instead of a monotonous decrease with increasing temperature. The unusual temperature dependence of MR has also been observed in the MBE $Mn_2CoAl$ films.[9] Considering that the negative MR originated from spin dependent scattering should enhance with lowering temperature, while nonstoichiometry-type impurity scatterings dominated at low temperatures reduce the MR effect, a maximum value was thus reached as a competitive result.

Our calculated density of states (DOS) for $Mn_2CoAl$ and $Mn_2Co_{0.8}Al_{1.2}$, respectively, is shown in Fig. 3 where the Fermi Level is set at zero. Notice that the Fermi level of $Mn_2Co_{0.8}Al_{1.2}$ moved away slightly from the valley thus destroying the gapless feature to some extent. Our calculation can be further supported by a recently paper,[15] where they also proposed that defects can destroy the SGS while lattice distortion would not.

The Hall resistivity $\rho_{xy}(H)$ curve in Fig. 4 present an anomalous Hall effect featured for magnetic systems with nonlinear and even hysteresis behavior at low fields. The overall Hall resistivity is generally expressed as $\rho_{xy}(H) = R_O H + R_A M(H)$, where $R_O$ and $R_A$ is the ordinary and anomalous Hall coefficient, respectively. Generally, the anomalous Hall resistivity can be scaled with the longitudinal resistivity in the form $\rho_{AHE} = a\rho_{xx} + b\rho_{xx}^2$, where $a$ contains information about skew scattering and $b$ relates to side jump mechanisms or intrinsic Berry phase contribution.[16,17] By extrapolating the high field $\rho_{xy}(H)$ to zero, we obtained $\rho_{AHE}$ for various temperatures and found a linear correlation of $\rho_{AHE} \sim \rho_{xx}^2$ as plotted in the upper inset, yielding a slope value of 5.3 S/cm. The anomalous Hall conductivity represented by $\sigma_{xy}^{AHE} = \rho_{AHE} / \rho_{xx}^2$, just equivalent to the slope, is actually a temperature independent value. We plot the Hall conductivity as a function of field at temperature of 5K in the lower inset of Fig.4, from which $\sigma_{xy}^{AHE}$ of ~5 S/cm is extracted, consistent with what the above slope determined. This is a smaller value than that of bulk $Mn_2CoAl$ [8] which is closer to the calculated value for $Mn_2CoAl$ based on Berry phase, thus favoring the intrinsic origin of the anomalous Hall effect. Moreover, based on the one band model $R_O = 1/nq$ and $\mu = R_O(T)/\rho(T)$, we obtain the carrier concentration and mobility to be $1.6\times10^{20}$ cm$^{-3}$ and 0.45 cm$^2$/V·s at 5K, respectively. The carrier mobility is comparable to the MBE $Mn_2CoAl$ films but is relatively lower than that of the bulk $Mn_2CoAl$, which is assumed to be resulted from impurity levels or atomic disorder scattering.

Figure 5(a) shows the magnetization curves of $Mn_2CoAl$ film measured at 5K with magnetic field applied in the film plane and out of plane, respectively. The saturation moment at 5K is determined to be 1.94 $\mu_B$ per formula unit. This value compares favorably with predicted theoretical value of 1.98 $\mu_B$, suggesting good crystalline of the film. The magnetization curve shows hysteresis behavior at low temperature, in accordance with the above Hall measurements result. The coercive filed at 5K is 1200Oe for in plane and 550Oe for out-of-plane configuration. Considering that bulk $Mn_2CoAl$ is a soft ferromagnet,[8, 18] the coercive field of the film should be attributed to shape anisotropy induced by film orientation.

Saturation magnetization as a function of temperature was measured over the range 5-400K. The saturation magnetization $M_s(T)$ curves are shown in Figure 5(b). The sample was first cooled down to 5 K and then warmed with a saturated magnetic field of 20 kOe applied parallel to the sample plane. We found that, for our $Mn_2CoAl$ film, the temperature dependence of the magnetization in saturation could be described, within reasonable limits and at the temperature range between 5 and 300K, by the Bloch formula: $M_s(T) = M_s(0)\left(1 - \mathrm{b}T^{3/2}\right)$, where $M_s(0)$ and b depend on the film thickness. As shown in the inset of Figure 4(b), fitting the data to this form yields a best fit value for b of $4.1\times10^{-5}$ $\mathrm{K}^{-3/2}$, Above 300K, our data for $Mn_2CoAl$ does not conform to the Bloch $T^{3/2}$ law; instead we found that M(T) can be fitted empirically as a function of temperature as $M_s(T) = M_s(0)[1-(T/T_c)^2]^{1/2}$. The Curie temperature ($T_c$) for the $Mn_2CoAl$ film can be estimated to be 550K. The resulting value is reduced by almost 200K comparing to that of the bulk $Mn_2CoAl$,[8] which is presumed to be caused by the deficiency of the magnetic atom Co based on the aforementioned composition analysis. Moreover, the magnetization change from $T^{3/2}$ to $T^2$ behavior has been previously observed in half-metallic Heusler alloys and can be attributed to a transition from localized to itinerant-like ferromagnetism.[19,20]

In summary, we have successfully grown oriented $Mn_2CoAl$ films on the thermally oxidized Si substrates by magnetron sputtering deposition. For seeking the signature of spin gapless semiconductor, we investigate in detail the magnetic and transport properties of the 200-nm-thickness film. As expected, the film reveals a semiconducting-like resistivity behavior and linear MR in the overall region. In addition, the MR presents a maximum as a competitive result between spin-dependent scattering and atomic disorder scattering for different temperatures. The unusual low anomalous Hall conductivity of our film is also an indication of its gapless band structure. The saturation magnetization of $Mn_2CoAl$ film approaches the theoretical value of 2.0$\mu_B$ while the Curie temperature (~550K) is lower than the bulk value of 720K due to Co deficiency. The carrier concentration ($1.6\times10^{20}cm^{-3}$) is approximately two orders magnitude higher than the reported bulk value. The results were discussed in term of the Fermi-level shift induced by the composition deviation. Thus, further experiments are necessary to control the composition and improve the ordering of the films.

## Acknowledgements

This work was supported by funding from the "973" Project (2012CB619405) and NSFC (Nos. 51171207, 51025103 and 11274371). We are grateful to Prof. R. Shan from Tongji University for fruitful discussions.

Figure captions:

FIG. 1. (Color online) XRD pattern of both the film (blue line) and substrate (grey line). All peaks are indexed except the asterisk (*) indicated ones that should belong to the silicon dioxides which is difficult to index due to their complex phases. The insets show the SEM image and the elemental content distribution scanned by EDX.

FIG. 2. (Color online) (a) Longitudinal resistivity as a function of temperature measured at zero magnetic field. The inset shows the fitting of conductivity. (b) Magnetoresistance (MR) measured with magnetic field perpendicular to the film plane, and the inset shows the absolute MR value variation with temperature.

FIG. 3. (Color online) Density of states of $Mn_2CoAl$ (blue line) and $Mn_2Co_{0.8}Al_{1.2}$ (red line).

FIG. 4. (Color online) Anomalous Hall effect measured for various temperatures. The upper inset plotted the anomalous Hall resistivity $\rho_{AHE}$ against the longitudinal resistivity $\rho_{xx}^2$ , with the red line representing a linear fit to it; The lower inset showed the Hall conductivity at 5K.

FIG. 5. (Color online) (a) Magnetization hysteresis loops of the synthesized film at 5K with the magnetic field applied in plane and perpendicular to the film, respectively. (b) Saturated magnetization $M_s(T)$ normalized by $M_s(0)$ plotted as function of temperature. The dotted red line represents the high temperature fit to the formula $Ms(T)=Ms(0)[1-(T/T_c)^2]^{1/2}$ , and the inset shows the fit to Bloch formula of $Ms(T)=Ms(0)(1-\mathrm{b}T^{3/2})$ below room temperature.

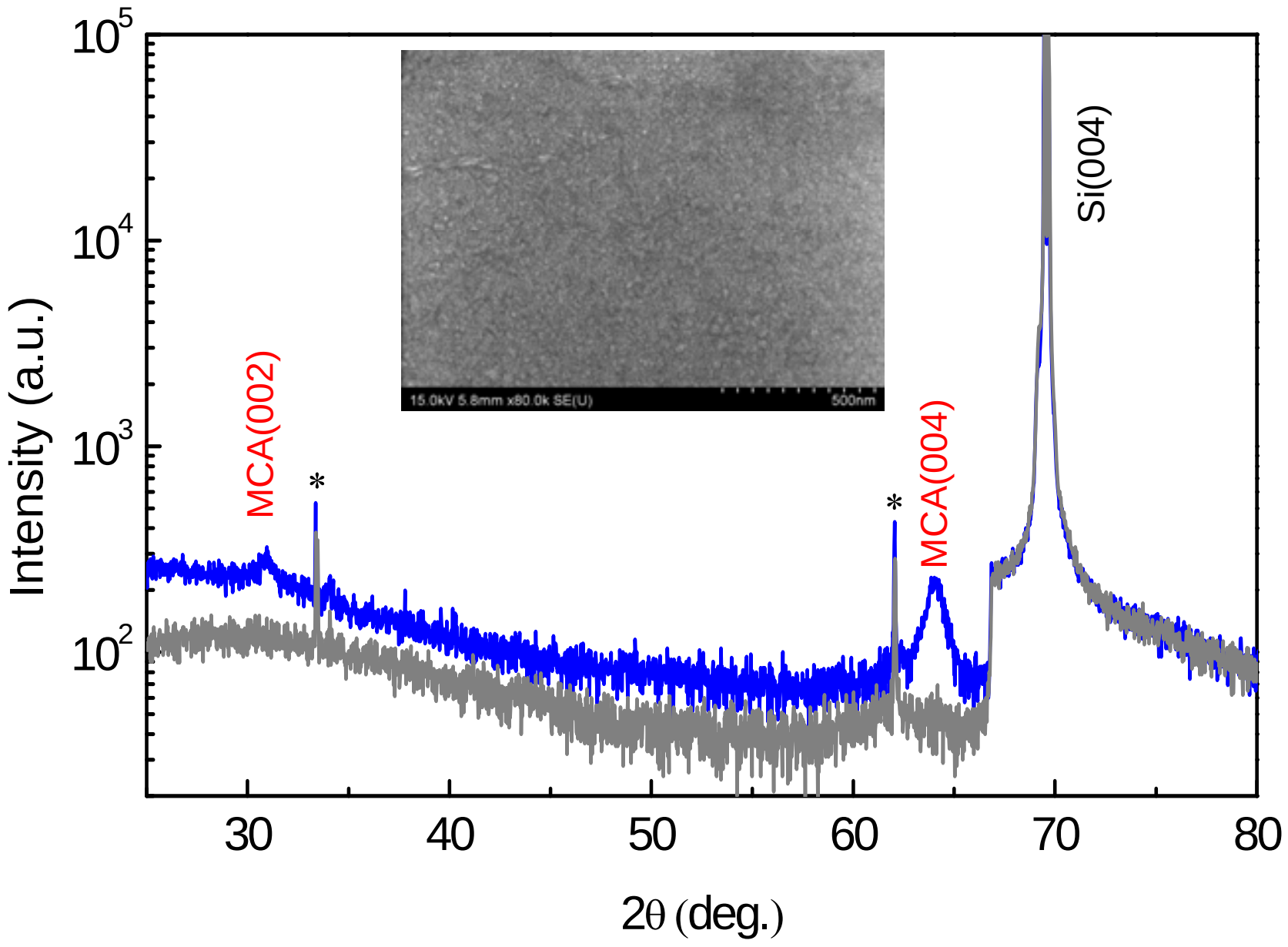

Figure 1 Xu et al.,

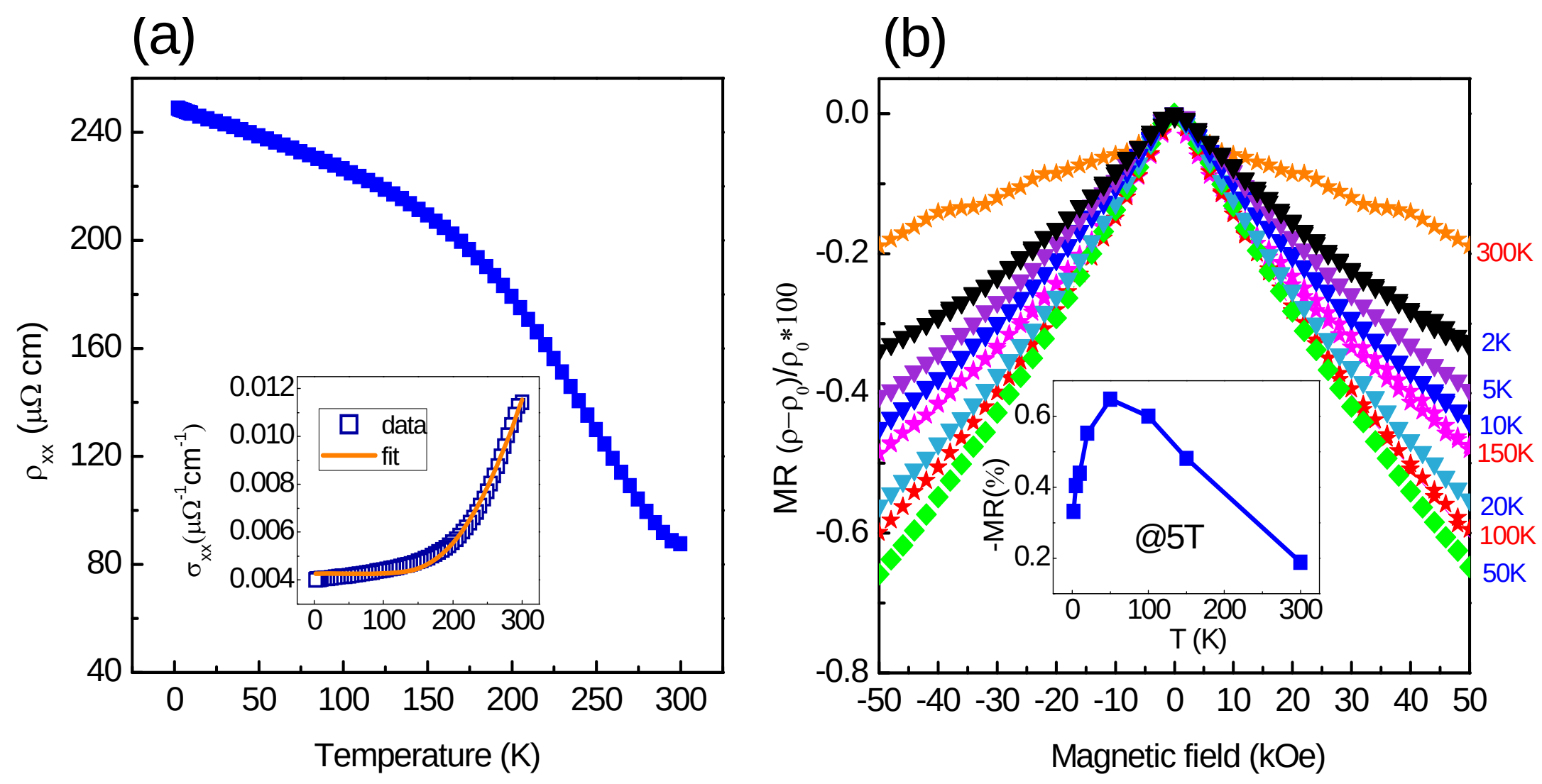

Figure 2 Xu et al.,

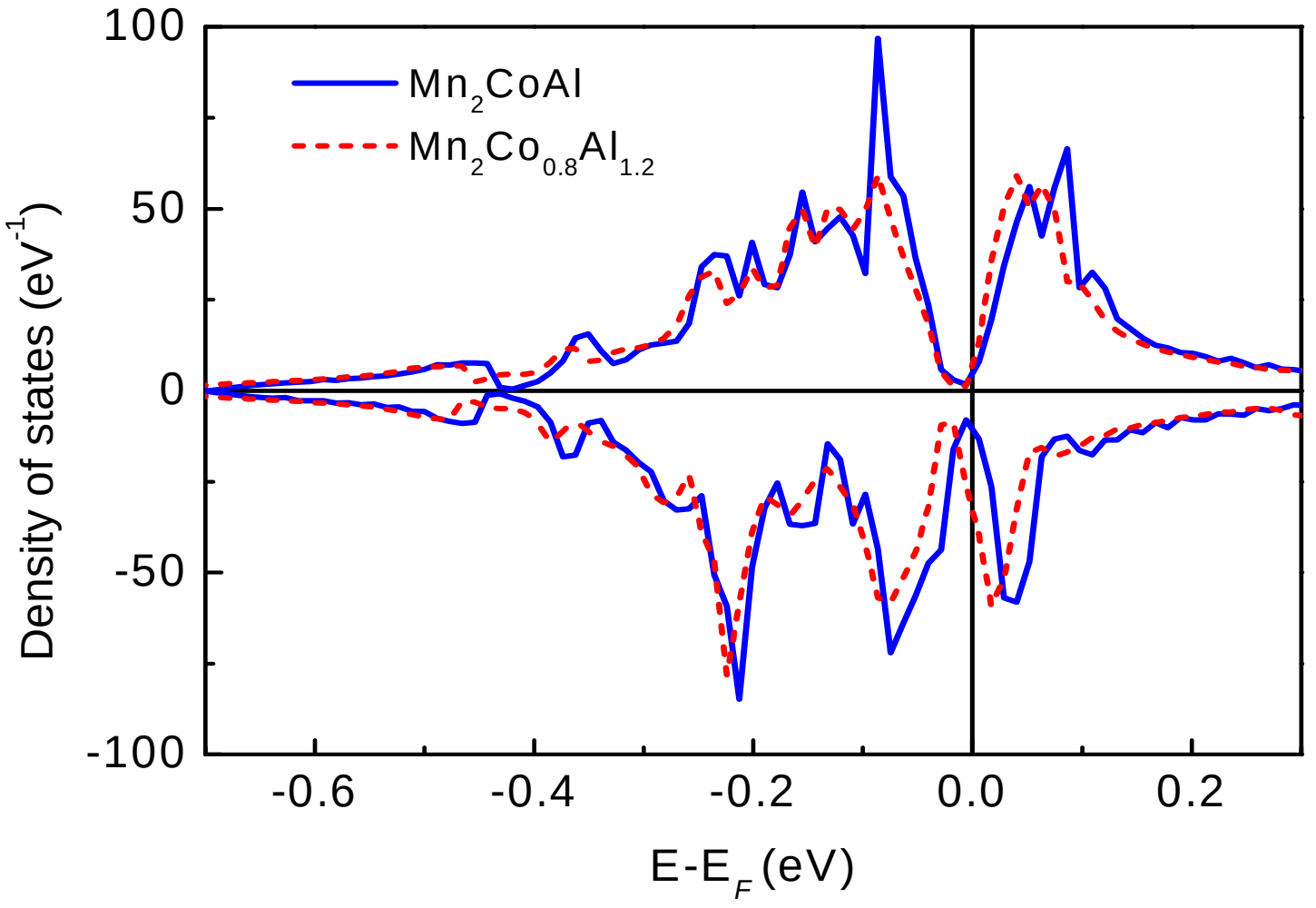

Figure 3 Xu et al.,

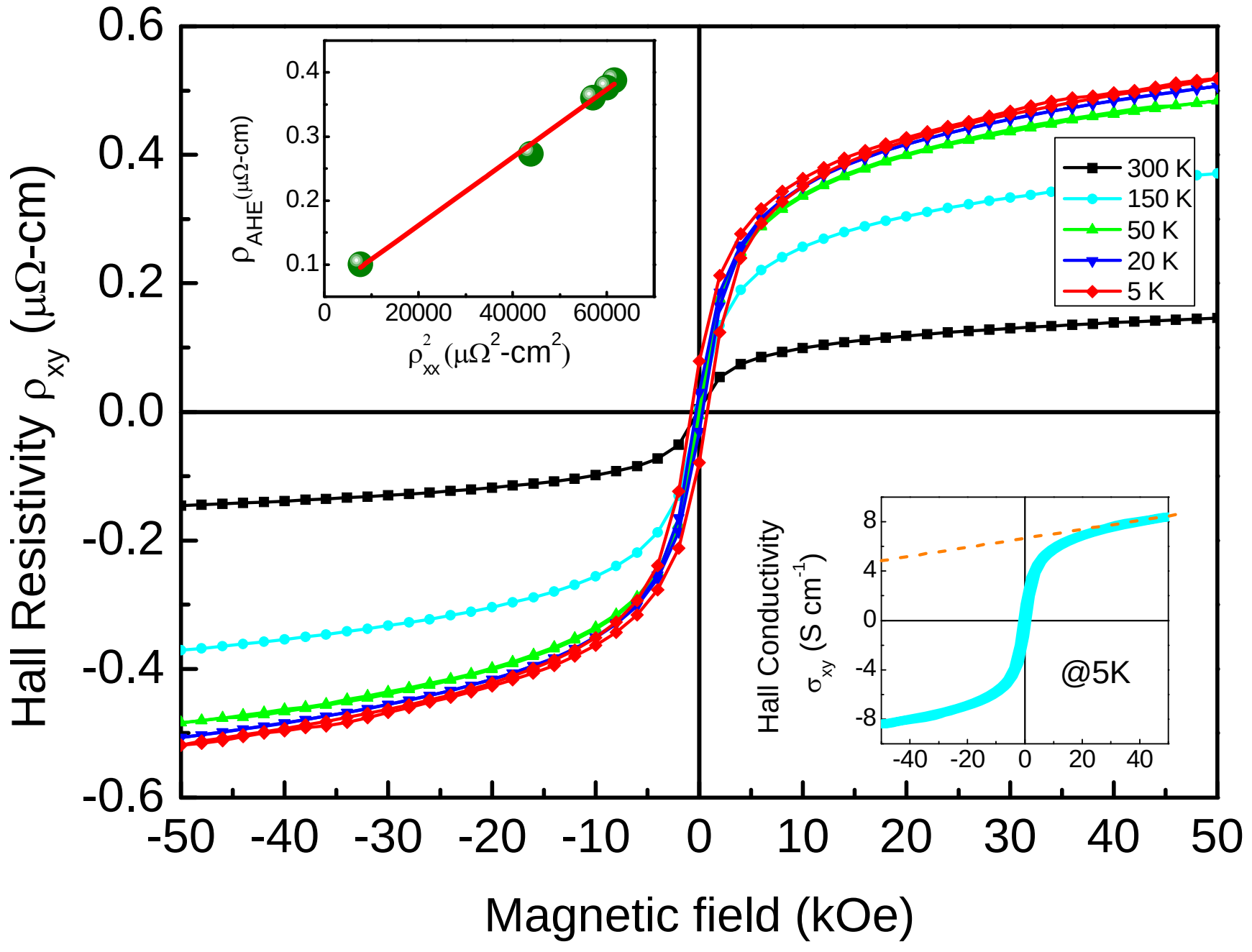

Figure 4 Xu et al.,

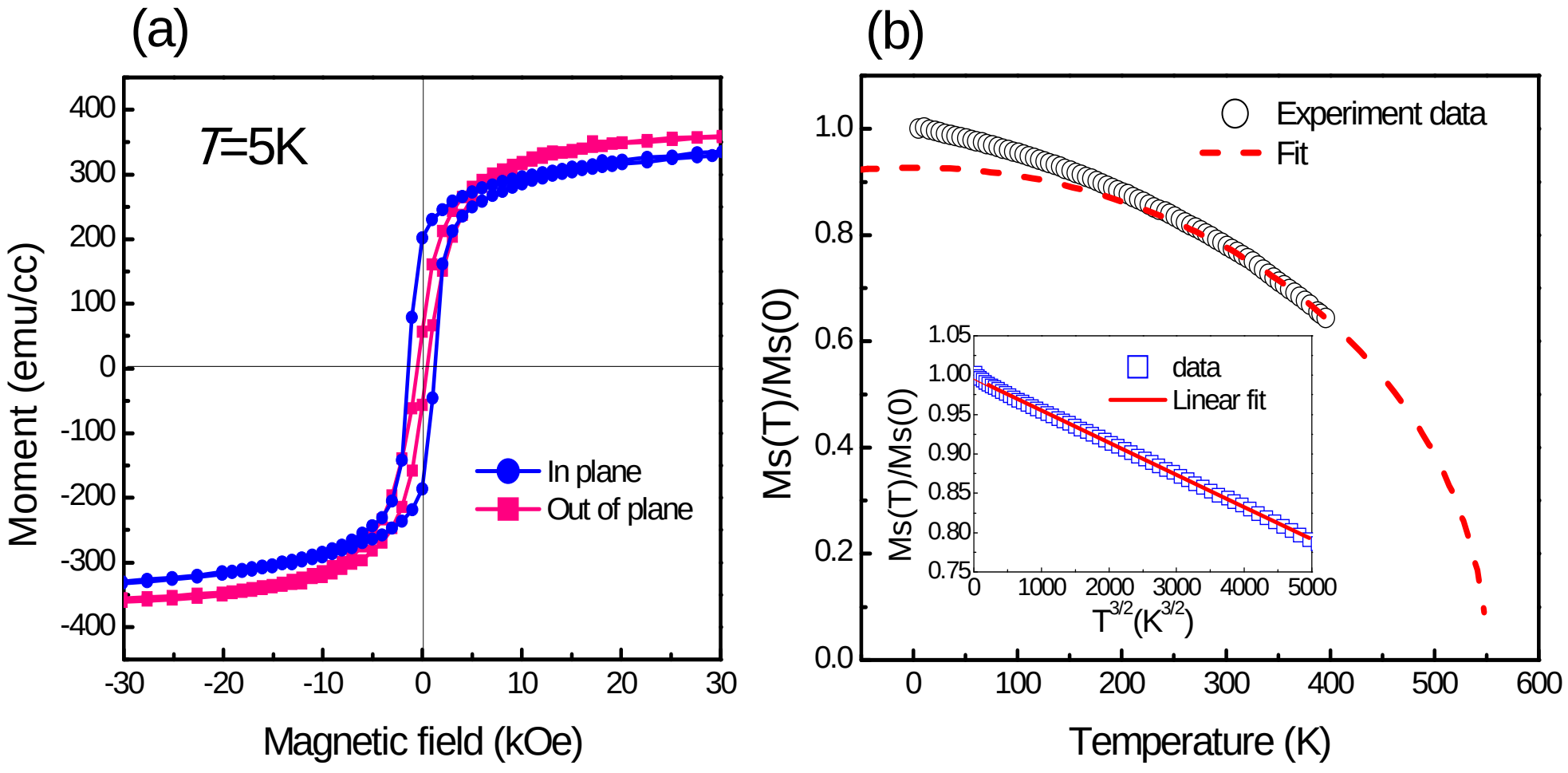

Figure 5 Xu et al.,